\documentclass{pasj00}
\draft

\begin{document}
\SetRunningHead{S. Kato}{Excitation of Disk Oscillations in Deformed Disks}
\Received{2008/0/00}
\Accepted{2008/07/25}

\title{Resonant Excitation of Disk Oscillations in Deformed Disks III:  
        Revision of Mathematical Treatment}

\author{Shoji \textsc{Kato}}%
\affil{2-2-2 Shikanodai-Nishi, Ikoma-shi, Nara 630-0114}
\email{kato.shoji@gmail.com; kato@kusastro.kyoto-u.ac.jp}


%


\maketitle

\begin{abstract}
In previous studies, we have examined a resonant excitation of disk oscillations 
in deformed disks.
In these studies, however, mathematical treatment around the resonant 
points was not rigorous.
In this paper the inadequate point is corrected, with no essential changes in the 
final results.
For this excitation process to work, disks must be general relativistic.
That is, the non-monotonic radial distribution  of epicyclic frequency in relativistic
disks is essential for the presence of the resonance and for trapping of oscillations.
In this paper, the growth rate of resonant oscillations is expressed in a form
more suitable for numerical calculations.
\end{abstract}


\section{Introduction}

In previous papers (Kato 2004, 2008a), we proposed a resonant excitation 
mechanism of disk oscillations in deformed disks.
The purpose is to propose an excitation mechanism of
quasi-periodic oscillations (QPOs) observed in neutron-star and 
black-hole X-ray binaries (e.g., Kato and Fukue 2007, Kato 2008b).
In this resonant excitation model of disk oscillations, a deformation
of disks from an axially-symmetric equilibrium state is essential.
The deformation to be considered is a warp or an eccentric deformation
of disks in the equatorial plane.

An outline of the model is as follows.
A non-linear coupling between a disk oscillation (hereafter we call it
the original oscillation) and a deformed part of the disks 
(warp or eccentric deformation) brings about some forced disk oscillations 
(we call them intermediate oscillations).
The intermediate oscillations make a resonant coupling with the unperturbed 
disk at particular radii of the disk.
After this resonant coupling, the intermediate oscillations feedback to the 
original oscillation by a nonlinear coupling with the deformed 
part of the disk.
Since this nonlinear feedback process involves a resonance, the original
oscillation is amplified or dampened.
Kato (2004, 2008a) examined this nonlinear feedback processes, and derived  
an excitation criterion and growth (damping) rate of the resonant oscillations.
It is of importance to note that in Keplerian disks this resonant excitation process
works only when the disks are general relativistic.
That is, a non-monotonic radial distribution of epicyclic frequency is necessary for 
appearance of resonance and for trapping of oscillations.

In these studies the intermediate disk oscillations were assumed to be local near to the
resonant radius in the sense that their radial wavelength is shorter than the 
characteristic radial length of disks.
That is, the radial derivative, $\partial /\partial r$, operated to the 
wave quantities was taken to be $ik$.
Further, $k$ was assumed to be constant.
This treatment of $k=$ const., however, was inadequate around the resonant point.
If this inadequate treatment is properly corrected, we find that unlike in the 
previous papers, the resonant point is not the place where a local dispersion 
relation of the intermediate oscillations is satisfied with a constant $k$, 
but the radii of Lindblad resonance for intermediate oscillations (when resonance occurs by
horizontal motions) or the radii where the intermediate oscillations are trapped in the 
vertical direction (when resonance occurs by vertical motions).

In this paper, we modify the analyses in our previous papers so that
the above-mentioned inadequate treatment is corrected.
We apply the mathematical techniques used by Meyer-Vernet and Sicardy (1987)
in their study of resonant disk-satellite interaction.
The results show that the stability criterion derived in the previous papers are 
unchanged.
The expression for growth rate obtained in the previous papers is found to be still 
applicable, but in this paper the expression is changed in a form more
suitable to numerical calculations.

We first summarize in section 2 the basic equations and relations
necessary to the study of the present resonant excitation problem,
although they are given in the previous papers.
In section 3, the stability criterion and growth rate are derived to the 
case of pressure-less disks as it is instructive,
and to the case of disks with pressure in section 4.
The final section is devoted to a brief discussion on a meaning of the instability criterion.

\section{Brief Summary of Basic Equations}

We briefly summarize here basic equations and relations [see Kato (2008a) for details]
that are necessary in the subsequent sections.

\subsection{Nonlinear Hydrodynamical Equations}

In the present problem, the general relativity is essential.
For simplicity, however, formulation in this paper is done in  
the framework of a pseudo-Newtonian using the 
gravitational potential introduced by Paczy{\' n}ski and Wiita (1980),
assuming that the central object has no rotation.
We adopt a Lagrangian formulation by 
Lynden-Bell and Ostriker (1967).

The unperturbed disk is in a steady state with a steady flow $\mbox{\boldmath $u$}_0$.
By using a displacement vector, $\mbox{\boldmath $\xi$}$, 
a weakly nonlinear hydrodynamical equation describing adiabatic, 
non-self-gravitating perturbations is written as, after generalizing the linear equation
derived by Lynden-Bell and Ostriker (1967),
\begin{equation}
    \rho_0{\partial^2\mbox{\boldmath $\xi$}\over\partial t^2}
         +2\rho_0(\mbox{\boldmath $u$}_0\cdot\nabla)
                 {\partial\mbox{\boldmath $\xi$}\over\partial t}
      +\mbox{\boldmath $L$}(\mbox{\boldmath $\xi$})
      =\rho_0\mbox{\boldmath $C$}(\mbox{\boldmath $\xi$},
                                                       \mbox{\boldmath $\xi$}),
\label{2.3}
\end{equation}
where $\mbox{\boldmath $L$}(\mbox{\boldmath $\xi$})$ is a linear
Hermitian operator with respect to $\mbox{\boldmath $\xi$}$ (Lynden-Bell and Ostriker 1967)
and is
\begin{eqnarray}
   \mbox{\boldmath $L$}(\mbox{\boldmath $\xi$})
     =\rho_0(\mbox{\boldmath $u$}_0\cdot\nabla)
                         (\mbox{\boldmath $u$}_0\cdot\nabla)\mbox{\boldmath $\xi$}
       +\rho_0(\mbox{\boldmath $\xi$}\cdot\nabla)(\nabla\psi_0)
       +\nabla\biggr[(1-\Gamma_1)p_0{\rm div}\mbox{\boldmath $\xi$}\biggr] 
                          \nonumber \\
       -p_0\nabla({\rm div}\mbox{\boldmath $\xi$})
       -\nabla[(\mbox{\boldmath $\xi$}\cdot\nabla)p_0]
       +(\mbox{\boldmath $\xi$}\cdot\nabla)(\nabla p_0),
\label{2.4}
\end{eqnarray}
and $\rho_0(\mbox{\boldmath $r$})$ and $p_0(\mbox{\boldmath $r$})$ are
the density and pressure in the unperturbed state, and $\Gamma_1$ is
the barotropic index specifying the linear part of the relation between 
Lagrangian variations $\delta p$ and $\delta \rho$, i.e.,
$(\delta p/p_0)_{\rm linear}=\Gamma_1(\delta\rho/\rho_0)_{\rm linear}$.
Since, the self-gravity of the disk gas has been neglected,
the gravitational potential, $\psi_0(\mbox{\boldmath $r$})$, is a given function and
there is no Eulerian perturbation of it. 
At the stage of equation (\ref{2.3}), there is no restriction on the form of 
$\mbox{\boldmath $u$}_0$, but in the main part of this paper, 
$\mbox{\boldmath $u$}_0$ is taken to be a pure cylindrical rotation.

The right-hand side of wave equation (\ref{2.3}) represents the 
weakly nonlinear terms.
Derivation of a detailed expression for $\mbox{\boldmath $C$}$ is boring, because of
the presence of many terms.
However, after lengthy manipulations we can summarize them into a relatively simple
form in the case of $\Gamma_1=1$:
\begin{equation}
    \rho_0\mbox{\boldmath $C(\xi, \xi)$}=
         -{1\over 2}\rho_0\xi_i\xi_j{\partial^2\over \partial r_i\partial r_j}(\nabla\psi_0)
         -{\partial\over\partial r_i}\biggr(p_0{\partial \xi_i\over\partial r_j}\nabla \xi_j\biggr),
\label{}
\end{equation}
using Cartesican cordinates.
When $\Gamma_1\not= 1$, some terms should be added on the right-hand side, which 
are shown in equation (82) of Kato (2008a).
An important characteristics of 
$\mbox{\boldmath $C$}$ is that we have commutative relations (Kato 2008a)
for an arbitrary set of $\mbox{\boldmath $\xi$}_1$, 
$\mbox{\boldmath $\xi$}_2$, and $\mbox{\boldmath $\xi$}_3$:
\begin{equation}
    \int\rho_0{\mbox{\boldmath $\xi$}}_1
                             \mbox{\boldmath $C$}
                          ({\mbox{\boldmath $\xi$}}_2, 
                           {\mbox{\boldmath $\xi$}}_3)dV
        =\int\rho_0{\mbox{\boldmath $\xi$}}_1
                             \mbox{\boldmath $C$}
                          ({\mbox{\boldmath $\xi$}}_3, 
                           {\mbox{\boldmath $\xi$}}_2)dV
        =\int\rho_0{\mbox{\boldmath $\xi$}}_3
                              \mbox{\boldmath $C$}
                           ({\mbox{\boldmath $\xi$}}_1, 
                            {\mbox{\boldmath $\xi$}}_2)dV.
\label{commutative}
\end{equation}
As shown later, the presence of these commutative relations is the reason why
we have a simple stability criterion.
We suppose that this commutative relation is a general property of conservative systems.

\subsection{Nonlinear Coupling and Growth Rate}

We now assume that the disks are deformed from an axisymmetric 
steady state by some external or internal cause.
The deformation is assumed, for simplicity, to be time-independent,
and has the azimuthal wavenumber $m_{\rm W}$.
The Lagrangian displacement associated with the deformation, 
$\mbox{\boldmath $\xi$}^{\rm W}(\mbox{\boldmath $r$}, t)$, is denoted by\footnote{
Here and hereafter, $\mbox{\boldmath $\xi$}^{\rm W}$, $\mbox{\boldmath $\xi$}$ 
[see equation (\ref{wave})],
and $\mbox{\boldmath $\xi$}^{\rm int}$ [see equation (\ref{intermediate})] 
are expressed in complex forms.
When their nonlinear couplings are calculated, we must be careful since the real displacement
associated with deformations, for example, is not $\mbox{\boldmath $\xi$}^{\rm W}$ itself,
but the real part of $\mbox{\boldmath $\xi$}^{\rm W}$ given here.
} 
\begin{equation}
      \mbox{\boldmath $\xi$}^{\rm W}(\mbox{\boldmath $r$}, t)
      ={\rm exp}(-im_{\rm W}\varphi)\hat{\mbox{\boldmath $\xi$}}^{\rm W}(r,z).
\label{warp}
\end{equation}
As the deformation, a warp will be the most probable, but it is not only one 
candidate of possible deformations.
A plane-symmetric one-armed spiral deformation is one of another possible 
candidates that can excite disk oscillations.
In both cases we have $m_{\rm W}=1$.

Our purpose here is to examine how the behavior of disk oscillations 
are affected by the disk deformation.
As shown below, some of oscillation modes are resonantly excited
on deformed disks through nonlinear coupling with disk deformation.
The nonlinear coupling processes are schematically shown in figure 1
of Kato (2004).

The displacement vector associated with a disk oscillation,
$\mbox{\boldmath $\xi$}$, is assumed to have frequency $\omega$ and
azimuthal wavenumber $m$.
Separating the time and azimuthal dependences from $\mbox{\boldmath $\xi$}
(\mbox{\boldmath $r$},t)$, we introduce $\hat {\mbox{\boldmath $\xi$}}$ as 
\begin{equation}
      \mbox{\boldmath $\xi$}(\mbox{\boldmath $r$},t)={\rm exp}[i(\omega t-m\varphi)]
      \hat {\mbox{\boldmath $\xi$}}(r,z).
\label{wave}
\end{equation}

The first step of the nonlinear interaction between the disk oscillation
characterized by $(\omega, m)$
and the deformation characterized by $(0,1)$ introduces two kinds 
of intermediate oscillations with azimuthal wavenumber being $m+1$ and $m-1$.
Let us denote the displacement vector associated with these intermediate
oscillations by 
\begin{equation}
    \mbox{\boldmath $\xi$}_\pm^{\rm int}(\mbox{\boldmath $r$},t)
        ={\rm exp}[i( \omega t-{\tilde m}\varphi)]
         {\hat{\mbox{\boldmath $\xi$}}}_\pm^{\rm int}(r,z),
\label{intermediate}
\end{equation}
where ${\tilde m}=m+1$ or $m-1$.
Here, ${\hat{\mbox{\boldmath $\xi$}}}_+^{\rm int}$ represents the intermediate 
oscillations resulting from the coupling between 
${\hat{\mbox{\boldmath $\xi$}}}$ and 
${\hat{\mbox{\boldmath $\xi$}}}^{\rm W}$, while 
${\hat{\mbox{\boldmath $\xi$}}}_-^{\rm int}$ does those resulting from
the coupling between ${\hat{\mbox{\boldmath $\xi$}}}$ and
${\hat{\mbox{\boldmath $\xi$}}}^{{\rm W}*}$, where the asterisk 
represents the complex conjugate.
That is, ${\hat{\mbox{\boldmath $\xi$}}}_+^{\rm int}$ and
${\hat{\mbox{\boldmath $\xi$}}}_-^{\rm int}$ are described, respectively, by
\begin{equation}
   -\omega^2\rho_0{\hat{\mbox{\boldmath $\xi$}}}_+^{\rm int}
  +2i\omega\rho_0(\mbox{\boldmath $u$}_0\cdot\nabla)
                      { \hat{\mbox{\boldmath $\xi$}}}_+^{\rm int}
  +\mbox{\boldmath $L$}({\hat{\mbox{\boldmath $\xi$}}}_+^{\rm int})
   ={1\over 2}[\rho_0\mbox{\boldmath $C$}({\hat{\mbox{\boldmath $\xi$}}},
                                  {\hat{\mbox{\boldmath $\xi$}}}^{\rm W})
                     +\rho_0\mbox{\boldmath $C$}
                                  ({\hat{\mbox{\boldmath $\xi$}}}^{\rm W},  
                                  {\hat{\mbox{\boldmath $\xi$}}})],
\label{2.7}
\end{equation}
\begin{equation}
   -\omega^2\rho_0{\hat{\mbox{\boldmath $\xi$}}}_-^{\rm int}
  +2i\omega\rho_0(\mbox{\boldmath $u$}_0\cdot\nabla)
                      { \hat{\mbox{\boldmath $\xi$}}}_-^{\rm int}
  +\mbox{\boldmath $L$}({\hat{\mbox{\boldmath $\xi$}}}_-^{\rm int})
        = {1\over 2}[\rho_0\mbox{\boldmath $C$}
                           ({\hat{\mbox{\boldmath $\xi$}}},
                                  {\hat{\mbox{\boldmath $\xi$}}}^{{\rm W}*})
                     +\rho_0\mbox{\boldmath $C$}
                                  ({\hat{\mbox{\boldmath $\xi$}}}^{{\rm W}*},  
                                   {\hat{\mbox{\boldmath $\xi$}}})].
\label{2.8}
\end{equation}

Next, the second stage of the nonlinear coupling is considered, which is a 
feedback process returning to the original oscillation, $\hat{\mbox{\boldmath $\xi$}}$,
by the intermediate oscillations $\hat{\mbox{\boldmath $\xi$}}^{\rm int}_\pm$ 
interacting with $\hat{\mbox{\boldmath $\xi$}}^{\rm W}$.
In the case where the coupling occurs through ${ \hat{\mbox{\boldmath $\xi$}}}_+^{\rm int}$,
the feedback is described by
\begin{equation}
   -\omega^2\rho_0{\hat{\mbox{\boldmath $\xi$}}}
  +2i\omega\rho_0(\mbox{\boldmath $u$}_0\cdot\nabla)
                      { \hat{\mbox{\boldmath $\xi$}}}
  +\mbox{\boldmath $L$}({\hat{\mbox{\boldmath $\xi$}}})
   ={1\over 2}[\rho_0\mbox{\boldmath $C$}
                   ({\hat{\mbox{\boldmath $\xi$}}}_+^{\rm int},
                    {\hat{\mbox{\boldmath $\xi$}}}^{{\rm W}*})
                    +\rho_0\mbox{\boldmath $C$}
                    ({\hat{\mbox{\boldmath $\xi$}}}^{{\rm W}*},  
                     {\hat{\mbox{\boldmath $\xi$}}}_+^{\rm int})].
\label{2.9}
\end{equation}
On the other hand, in the case where the feedback occurs through 
${\hat{\mbox{\boldmath $\xi$}}}_-^{\rm int}$,
the equation corresponding to equation (\ref{2.9}) is 
\begin{equation}
       -\omega^2\rho_0{\hat{\mbox{\boldmath $\xi$}}}
           +2i\omega\rho_0(\mbox{\boldmath $u$}_0\cdot\nabla)
                      { \hat{\mbox{\boldmath $\xi$}}}
           +\mbox{\boldmath $L$}({\hat{\mbox{\boldmath $\xi$}}})
       = {1\over 2}[\rho_0\mbox{\boldmath $C$}
                   ({\hat{\mbox{\boldmath $\xi$}}}_-^{\rm int},
                    {\hat{\mbox{\boldmath $\xi$}}}^{{\rm W}})
                    +\rho_0\mbox{\boldmath $C$}
                    ({\hat{\mbox{\boldmath $\xi$}}}^{{\rm W}},  
                     {\hat{\mbox{\boldmath $\xi$}}}_-^{\rm int})].
\label{2.10}
\end{equation}

An important point to be noted here is that as a result of this feedback
process, the original disk oscillation is amplified or dampened, i.e.,
the frequency, $\omega$, can be no longer real, since in the feedback processes
a resonance is involved.
How much is the imaginary part of $\omega$?
This can be examined from equations (\ref{2.9}) and (\ref{2.10}) 
using the fact that the operators $i\rho_0(\mbox{\boldmath{$u$}}_0\cdot\nabla)$ and  
$\mbox{\boldmath $L$}$ are Hermitian (Lynden-Bell and Ostriker 1967).
After some calculations the imaginary part of frequency, $\omega_{\rm i}$,
can be formally expressed as (e.g., Kato 2008a)
\begin{equation}
     -\omega_{{\rm i}, \pm} ={W_\pm \over 2E},
\label{growth}
\end{equation}
where $\pm$ denotes the cases of the coupling through ${ {\hat{\mbox{\boldmath $\xi$}}}}_+$
and ${{\hat{\mbox{\boldmath $\xi$}}}}_-$, respectively,
and
\begin{equation}
    W_+={\omega_0\over 2}\Im\int {1\over 2}\rho_0{\hat{\mbox{\boldmath $\xi$}}}^*
       [\mbox{\boldmath $C$}({\hat{\mbox{\boldmath $\xi$}}}_+^{\rm int},
                        {\hat{\mbox{\boldmath $\xi$}}}^{{\rm W}*})
     + \mbox{\boldmath $C$}({\hat{\mbox{\boldmath $\xi$}}}^{{\rm W}*},
                        {\hat{\mbox{\boldmath $\xi$}}}_+^{\rm int})] dV,
\label{2.16}
\end{equation}
\begin{equation}
    W_-={\omega_0\over 2} \Im\int {1\over 2}\rho_0 {\hat{\mbox{\boldmath $\xi$}}}^*
       [\mbox{\boldmath $C$}({\hat{\mbox{\boldmath $\xi$}}}_-^{\rm int},
                        {\hat{\mbox{\boldmath $\xi$}}}^{{\rm W}})
     + \mbox{\boldmath $C$}({\hat{\mbox{\boldmath $\xi$}}}^{{\rm W}},
                        {\hat{\mbox{\boldmath $\xi$}}}_-^{\rm int})] dV,
\label{2.18}
\end{equation}
and 
\begin{equation}
       E={1\over 2}\omega_0\int\rho_0{\hat{\mbox{\boldmath $\xi$}}}^*
               [\omega-i(\mbox{\boldmath $u$}\cdot\nabla)]
                        {\hat{\mbox{\boldmath $\xi$}}}dV,
\label{2.12}
\end{equation}
where $\omega_0$ is the frequency of the original oscillation before the mode couplings.
The above expressions for $W_\pm$ and $E$ have physical meanings that $W_\pm$ is the rate 
at which work is done on the original oscillations by the nonlinear resonant processes,
and $E$ is the wave energy of the original oscillations [see equation (93) by Kato (2001)].

It is important to change the above expressions for $W_\pm$ by using the
commutative relation (\ref{commutative}) as
\begin{equation}
   W_+={\omega_0\over 2}\Im \int{1\over 2}\rho_0{\hat{\mbox{\boldmath $\xi$}}}_+^{\rm int}
         [\mbox{\boldmath $C$}({\hat{\mbox{\boldmath $\xi$}}}^*,
                  {\hat{\mbox{\boldmath $\xi$}}}^{{\rm W}*})
        +\mbox{\boldmath $C$}({\hat{\mbox{\boldmath $\xi$}}}^{{\rm W}*},
                  {\hat{\mbox{\boldmath $\xi$}}}^*)]dV,
\label{work+}
\end{equation}
\begin{equation}
     W_-={\omega_0\over 2}\Im \int{1\over 2}\rho_0{\hat{\mbox{\boldmath $\xi$}}}_-^{\rm int}
         [\mbox{\boldmath $C$}({\hat{\mbox{\boldmath $\xi$}}}^*,
                  {\hat{\mbox{\boldmath $\xi$}}}^{{\rm W}})
        +\mbox{\boldmath $C$}({\hat{\mbox{\boldmath $\xi$}}}^{{\rm W}},
                  {\hat{\mbox{\boldmath $\xi$}}}^*)]dV.
\label{work-}
\end{equation}

\subsection{Oscillations in Isothermal Disks}

Hereafter, we restrict our attention to oscillations in geometrically 
thin disks.
The steady unperturbed disks are axially-symmetric and
have no motion except for rotation.
Here, cylindrical coordinates ($r$, $\varphi$, $z$) are employed, in which 
the $z$-axis is perpendicular to the disk plane and the origin of the
coordinates is at the disk center.
The unperturbed flow is then described as $\mbox{\boldmath $u$}_0=(0,r\Omega(r),0)$,
where $\Omega(r)$ is the angular velocity of disk rotation.
We further assume, for simplicity, that the disk is isothermal in the vertical 
direction.
In the vertically isothermal disks, 
the unperturbed density in disks, $\rho_0(r,z)$, is stratified as (e.g., Kato et al. 1998)
\begin{equation}
         \rho_0(r,z)=\rho_{00}(r){\rm exp}\biggr[-{z^2\over 2H^2(r)}\biggr],
\label{rho}
\end{equation}
where $\rho_{00}$ is the density on the equatorial plane, and $H$
is the half-thickness of the disk and  is related to the vertical epicyclic frequency,
$\Omega_\bot$,  by
\begin{equation}
       \Omega_\bot^2H^2={p_0\over \rho_0}=c^2_{\rm s}(r), 
\label{sound}
\end{equation}
where $c_{\rm s}$ is the isothermal acoustic speed.
The vertical epicyclic frequency, $\Omega_\bot$, is equal to the angular velocity
of the Keplerian rotation, $\Omega_{\rm K}$, in the case of the central object 
being non-rotating, and practically equal to $\Omega$, since the disk is assumed to be 
geometrically thin.
Hereafter, however, we use $\Omega_\bot$ without using $\Omega_{\rm K}$ or
$\Omega$, so that we can trace back the effects of $\Omega_\bot$ on the final
results.  

Concerning oscillations superposed on such disks, we consider oscillations whose 
radial wavelength is moderately short  
so that the radial variations of physical quantities in the unperturbed state 
[including the radial variation of $H(r)$] are
neglected compared with the radial variation of wave quantities.
Furthermore, we assume that the oscillations also occur isothermally.   
Then, we can neglect the term of $\nabla[(1-\Gamma_1)p_0{\rm div}
\mbox{\boldmath $\xi$}]$ in the operator $\mbox{\boldmath $L$}$ in equation (\ref{2.4}).

If the above simplifications and approximations are adopted,
the $r$- and $z$- dependences of $\hat{\mbox{\boldmath $\xi$}}(r,z)$ 
are approximately separated. 
That is, the $z$-dependence of 
$\hat{\mbox{\boldmath $\xi$}}(r,z)$ is obtained by solving an
eigen-value problem, and the eigen-functions are found to be an orthogonal set of 
Hermite polynomials (Okazaki et al. 1987).
Restricting our attention to one of such oscillations, we write
$\hat{\mbox{\boldmath $\xi$}}(r,z)$ as
\begin{equation}
      \hat{\xi}_r (r,z)=\breve{\xi}_{r,n}(r){\cal H}_n(z/H)                        
\label{xi-r}
\end{equation}
\begin{equation}
      \hat{\xi}_\varphi (r,z) 
           = \breve{\xi}_{\varphi,n}(r){\cal H}_n(z/H),
\label{xi-varphi}
\end{equation}
\begin{equation}
      \hat{\xi}_z (r,z)=\breve{\xi}_{z, n}(r){\cal H}_{n-1}(z/H).
\label{xi-z}
\end{equation}
where ${\cal H}_n$ is the Hermite polynomial of argument $z/H$, and
$n(=0,1,2,...)$ characterizes the number of node(s) of oscillations
in the vertical direction.
It is noted that the number of node(s) of $\hat{\xi}_z$ in the vertical
direction is smaller than those of $\hat{\xi}_r$ and $\hat{\xi}_\varphi$
by one, as shown in equation (\ref{xi-z}).
However, the subscript $n$ (not $n-1$) is attached to $\breve{\xi}_z$ 
as $\breve{\xi}_{z,n}$ in order to emphasize that $\breve{\xi}_{r,n}$, 
$\breve{\xi}_{\varphi,n}$,  and $\breve{\xi}_{z, n}$ are a set of
solutions.

Under these approximations, we express the $r$-, $\varphi$-, and
$z$- components of the homogeneous parts of wave equation (\ref{2.3}) as
\begin{equation}
     \biggr[-(\omega-m\Omega)^2+\kappa^2-4\Omega^2
         -c_{\rm s}^2{d^2\over dr^2}\biggr]\breve{\xi}_{r,n}
         -i2\Omega(\omega-m\Omega)\breve{\xi}_{\varphi,n}
         +\Omega^2_\bot H{d \breve{\xi}_{z,n}\over dr}
                =0,
\label{4.1}
\end{equation}
\begin{equation}
      -(\omega-m\Omega)^2\breve{\xi}_\varphi
              +i2\Omega(\omega-m\Omega)\breve{\xi}_r=0,
\label{4.2}
\end{equation}
\begin{equation}
     [-(\omega-m\Omega)^2+n\Omega^2_\bot]\breve{\xi}_{z,n}
            -n\Omega^2_\bot H{d \breve{\xi}_{r,n}\over dr}
        =0.
\label{4.3}
\end{equation}

Nonlinear coupling between ${\mbox{\boldmath $\xi$}}$ and
${\mbox{\boldmath $\xi$}}^{\rm W}$ introduce intermediate 
oscillations ${\mbox{\boldmath $\xi$}}^{\rm int}$.
The azimuthal wavenumber of the intermediate oscillations are 
$m+1$ or $m-1$.
Their $z$-dependence are characterized by
$n+1$ or $n-1$ when $n_{\rm W}=1$ since ${\cal H}_1{\cal H}_n={\cal H}_{n+1}
+n{\cal H}_{n-1}$ and by $n$ when $n_{\rm W}=0$.
To consider these various coupling cases separately, we write 
$\hat{{\mbox{\boldmath $\xi$}}}^{\rm int}_\pm$ in the forms of 
\begin{equation}
     \hat{\xi}_{r,\pm}^{\rm int}(r,z)
          =\breve{\xi}^{\rm int}_{r,\pm,\tilde{n}}(r){\cal H}_{\tilde{n}}(z/H),
\label{5.4}
\end{equation}
\begin{equation}
     \hat{\xi}_{\varphi,\pm}^{\rm int}(r,z)
         =\breve{\xi}^{\rm int}_{\varphi,\pm,\tilde{n}}(r) 
                       {\cal H}_{\tilde{n}}(z/H),
\label{5.5}
\end{equation}
\begin{equation}
     \hat{\xi}_{z,\pm}^{\rm int}(r,z)
            =\breve{\xi}^{\rm int}_{z,\pm,\tilde{n}} (r)
                    {\cal H}_{\tilde{n}-1}(z/H),
\label{5.6}
\end{equation}
where the subscript $\tilde{n}$ represents $n+1$ or $n-1$ or $n$.

Nonlinear coupling terms are also separated into terms proportional
to exp$[-i(m\pm 1)\varphi]$ and ${\cal H}_{\tilde n}(r/z)$.
That is, in the case of coupling through $\mbox{\boldmath $\xi$}_+^{\rm int}$,
we write the coupling terms as
\begin{equation}
      {1\over 2}\rho_0[ \mbox{\boldmath $C$}({\mbox{\boldmath $\xi$}},
                     {\mbox{\boldmath $\xi$}}^{\rm W})
              + \mbox{\boldmath $C$}({\mbox{\boldmath $\xi$}}^{\rm W},
                      {\mbox{\boldmath $\xi$}})]_r
       =\rho_0\sum_{{\tilde m},\tilde{n}} \breve{A}_{r,+,\tilde{n}}(r)
               {\rm exp}[i(\omega t-{\tilde m}\varphi)]
                {\cal H}_{\tilde {n}}(z/H)+\cdot\cdot\cdot
\label{5.7}
\end{equation}
\begin{equation}
      {1\over 2}\rho_0[ \mbox{\boldmath $C$}({\mbox{\boldmath $\xi$}},
                     {\mbox{\boldmath $\xi$}}^{\rm W})
              + \mbox{\boldmath $C$}({\mbox{\boldmath $\xi$}}^{\rm W},
                      {\mbox{\boldmath $\xi$}})]_\varphi
       =\rho_0\sum_{{\tilde m},\tilde{n}} \breve{A}_{\varphi,+,\tilde{n}}(r)
                {\rm exp}[i(\omega t-{\tilde m}\varphi)]
                {\cal H}_{\tilde {n}}(z/H)+\cdot\cdot\cdot
\label{5.7'}
\end{equation}
\begin{equation}
      {1\over 2}\rho_0[ \mbox{\boldmath $C$}({\mbox{\boldmath $\xi$}},
                     {\mbox{\boldmath $\xi$}}^{\rm W})
              + \mbox{\boldmath $C$}({\mbox{\boldmath $\xi$}}^{\rm W},
                      {\mbox{\boldmath $\xi$}})]_z
       =\rho_0\sum_{{\tilde m},\tilde{n}} \breve{A}_{z,+,\tilde{n}}(r)
                {\rm exp}[i(\omega t-{\tilde m}\varphi)]
                {\cal H}_{\tilde {n}-1}(z/H)+\cdot\cdot\cdot,
\label{5.7''}
\end{equation}
where $+\cdot\cdot\cdot$ denotes terms orthogonal both to 
${\cal H}_{\tilde{n}}$ and
${\cal H}_{\tilde{n}-1}$, and the subscript $+$ is added to $\breve{A}$'s 
in order to emphasize that they are related to the $\varphi$-dependence of 
${\rm exp}[-i(m+1)\varphi]$.
In a similar way, in the case of coupling through the intermediate oscillations
of ${\mbox{\boldmath $\xi$}}_-^{\rm int}$,
$(1/2)\rho_0[ \mbox{\boldmath $C$}({\mbox{\boldmath $\xi$}},
{\mbox{\boldmath $\xi$}}^{{\rm W}*})+
\mbox{\boldmath $C$}({\mbox{\boldmath $\xi$}}^{{\rm W}*},
{\mbox{\boldmath $\xi$}})]$ can be expressed in forms similar to 
equations (\ref{5.7}) -- (\ref{5.7''}), introducing 
$\breve{A}_{r,-,\tilde{n}}$, $\breve{A}_{\varphi,-,\tilde{n}}$, 
$\breve{A}_{z,-,\tilde{n}}$.

The equations describing intermediate oscillations are then written as
\begin{eqnarray}
     \biggr[-(\omega-{\tilde m}\Omega)^2+\kappa^2-4\Omega^2
         -c_{\rm s}^2{d^2\over dr^2}\biggr]
             \breve{\xi}^{\rm int}_{r,\pm,{\tilde n}}
         -i2\Omega(\omega-{\tilde m}\Omega)
             \breve{\xi}^{\rm int}_{\varphi,\pm,{\tilde n}}  \nonumber \\
         +\Omega^2_\bot H{\partial \breve{\xi}^{\rm int}_{z,\pm,{\tilde n}}
               \over\partial r}
                =\breve{A}_{r,\pm,\tilde{n}},
\label{4.1'}
\end{eqnarray}
\begin{equation}
      -(\omega-{\tilde m}\Omega)^2\breve{\xi}^{\rm int}_{\varphi,\pm,{\tilde n}}
              +i2\Omega(\omega-{\tilde m}\Omega)\breve{\xi}^{\rm int}_{r,\pm,{\tilde n}}
                 =\breve{A}_{\varphi, \pm,\tilde{n}},
\label{4.2'}
\end{equation}
\begin{equation}
     [-(\omega-{\tilde m}\Omega)^2+{\tilde n}\Omega^2_\bot]
            \breve{\xi}^{\rm int}_{z,\pm, {\tilde n}}
            -{\tilde n}\Omega^2_\bot H{d 
            \breve{\xi}^{\rm int}_{r,\pm,{\tilde n}}\over dr}
        =\breve{A}_{z, \pm,\tilde{n}}, 
\label{4.3'}
\end{equation}
where ${\tilde m}$ represents $m+1$ or $m-1$, and ${\tilde n}$ does
$n+1$ or $n-1$ when $n_{\rm W}=1$ and ${\tilde n}=n$ when $n_{\rm W}=0$. 
From equations (\ref{4.1'}) and (\ref{4.2'}) we can eliminate 
${\breve \xi}^{\rm int}_{\varphi,\pm, {\tilde n}}$ to give 
\begin{equation}
     [-(\omega-{\tilde m}\Omega)^2+\kappa^2]\breve{\xi}^{\rm int}_{r,\pm, {\tilde n}}
       -c_{\rm s}^2{d^2 \breve{\xi}^{\rm int}_{r,\pm, {\tilde n}}
              \over dr^2}
       +\Omega^2_\bot H {d\breve{\xi}^{\rm int}_{z,\pm, {\tilde n}}
            \over dr}
        =\breve{A}_{r,\pm,{\tilde n}}
         -i{2\Omega\over\omega-{\tilde m}\Omega}\breve{A}_{\varphi,\pm,{\tilde n}}.
\label{4.1''}
\end{equation}
Hereafter, we use equations (\ref{4.1'}) -- (\ref{4.1''}) as the basic
equations describing the intermediate oscillations.

Finally, we should notice that in the case of isothermal disks described above, 
$W_+$ and $W_-$ given by equations (\ref{work+}) and (\ref{work-}) are written in the forms of
\begin{equation}
   W_{+}=\frac{\omega_0}{2}\Im \int\rho_{00}(x)[(2\pi)^{3/2}{\tilde n}!rH] 
       \cdot \biggr[{\breve \xi}_{r,+}^{\rm int}{\breve A}_{r,+}^*
              +{\breve \xi}_{\varphi,+}^{\rm int}{\breve A}_{\varphi,+}^*
              +\frac{1}{{\tilde n}}{\breve \xi}_{z, +}{\breve A}_{z,+}^*\biggr]dx,
\label{W+}
\end{equation}
and
\begin{equation}
   W_{-}=\frac{\omega_0}{2}\Im \int\rho_{00}(x)[(2\pi)^{3/2}{\tilde n}!rH] 
       \cdot \biggr[{\breve \xi}_{r,-}^{\rm int}{\breve A}_{r,-}^*
              +{\breve \xi}_{\varphi,-}^{\rm int}{\breve A}_{\varphi,-}^*
              +\frac{1}{{\tilde n}}{\breve \xi}_{z, -}{\breve A}_{z,-}^*\biggr]dx,
\label{W-}
\end{equation}
where $[(2\pi)^{3/2}{\tilde n}!rH]$ comes from the part of volume integration
in the $\varphi$- and $z$-directions.
Furthermore, 
the wave energy, $E$, given by equation (\ref{2.12}) can be expressed as
\begin{equation}
    E={(2\pi)^{3/2}\over 2}\omega_0^2(r^4H\rho_{00})_{\rm c}E_n,
\label{5.28}
\end{equation}
where $E_n$ is a dimensionless quantity given by\footnote{
This expression for $E_n$ is different from that given in the previous paper (Kato 2008a),
since the previous one is not suitable to represent the sign of wave energy.
}
\begin{equation}
    E_n=\int {rH\rho_{00}\over (rH\rho_{00})_{\rm c}}{\omega-m\Omega\over \omega}
        \biggr(n!{\vert \breve{\xi}_r\vert ^2\over r_{\rm c}^2}
        +(n-1)!{\vert \breve{\xi}_z\vert ^2\over r_{\rm c}^2}\biggr){dr\over r_{\rm c}},
\label{5.29}
\end{equation}
and the subscript c denotes the values at resonant radius, which is defined by $J_1=0$
or $J_2=0$ (see below).
It is noted that in the case of $n=0$, $(n-1)!$ is zero.

\section{Resonant Excitation of Oscillations in Pressure-less Disks}

Before examining general cases of disks with pressure, we study here the limiting
case of pressure-less disks, i.e., $c_{\rm s}=0$ and $H=0$.
Equations (\ref{4.1''}) and (\ref{4.3'}) show that $\breve{\xi}_r^{\rm int}$ 
(and $\breve{\xi}_\varphi^{\rm int}$) and
$\breve{\xi}_z^{\rm int}$ respond resonantly, respectively, at the radii where
$-(\omega-{\tilde m}\Omega)^2+\kappa^2=0$ holds and at the radii where
$-(\omega-{\tilde m}\Omega)^2+{\tilde n}\Omega_\bot^2=0$ holds.
We define here $J_1(r)$ and $J_2(r)$ by
\begin{equation}
      J_1(r)=-(\omega-{\tilde m}\Omega)^2+\kappa^2 \quad {\rm and} \quad
      J_2(r)=-(\omega-{\tilde m}\Omega)^2+{\tilde n}\Omega_\bot^2.
\label{eqs-J}
\end{equation}
We call the resonance at $J_1=0$ horizontal resonances (Lindblad resonances) and 
the resonance at $J_2=0$ vertical resonances.
We consider these two resonances separately.

As a typical example in infinitesimally thin disks,
we consider here the case where all motions including those associated with disk
deformations are in the equatorial plane, i.e., $n=0$, $n_{\rm W}=0$, and
${\tilde n}=0$.\footnote{
In the following formulations, however, we retain $n$, $n_{\rm W}$, and ${\tilde n}$
in general without specifying to particular values.
}
At the radii of $J_1=0$, 
$\breve{\xi}_r^{\rm int}$ and $\breve{\xi}_\varphi^{\rm int}$ resonantly
respond to $\breve{A}$'s as [see equations (\ref{4.1''}) and (\ref{4.2'})]
\begin{equation}
    \breve{\xi}_{r,\pm,{\tilde n}}^{\rm int}={1\over J_1}
          \biggr(\breve{A}_{r,\pm,{\tilde n}}-i{2\Omega\over \omega-{\tilde m}\Omega}
            \breve{A}_{\varphi,\pm,{\tilde n}}\biggr),
\label{3.1}
\end{equation}
\begin{equation}
    \breve{\xi}_{\varphi,\pm,{\tilde n}}^{\rm int}={1\over J_1}
       \biggr[ i{2\Omega\over\omega-{\tilde m}\Omega}\breve{A}_{r,\pm,{\tilde n}}
          -{-(\omega-{\tilde m}\Omega)^2+\kappa^2-4\Omega^2\over 
          (\omega-{\tilde m}\Omega)^2}
           \breve{A}_{\varphi,\pm,{\tilde n}}\biggr].
\label{3.2}
\end{equation}
The quantity $\breve{\xi}_z^{\rm int}$ can be taken to be zero, since it does not
respond resonantly and has no contribution to the final results.

To examine the resonant interaction at the radii of $J_1=0$, 
we introduce a small imaginary part of
$\omega$, i.e., $\omega_{\rm i}$, as $\omega=\omega_{\rm r}+i\omega_{\rm i}$, 
where the imaginary part, $\omega_{\rm i}$, is
tentatively assumed to be negative so that causality is satisfied.
Later, the results obtained by $\omega_{\rm i}<0$ is extended analytically to the
whole region of $\omega_{\rm i}$ as usually done in stability analyses.
Hereafter, the real part, $\omega_{\rm r}$, is written by $\omega$ or $\omega_0$, without confusion.

Near to the resonance, $J_1$ is expanded as
\begin{equation}
   J_1=J_1'\biggr[(r-r_{\rm c}) 
       -i{2(\omega-{\tilde m}\Omega)_{\rm c}\over J_1'}\omega_{\rm i}\biggr],
\label{3.4}
\end{equation}
where the subscript c represents the values at the resonant radii, and $J_1'$ is
\begin{equation}
    J_1'=2\biggr[{\tilde m}(\omega-{\tilde m}\Omega)\Omega{d{\rm ln}\Omega\over dr}
         +\kappa^2{d{\rm ln}\kappa\over dr}\biggr]_{\rm c}.
\label{3.5}
\end{equation}
Let us now consider an integration of $f(r)/J_1$, $f$ being an arbitrary smooth
real function, along the radial direction including the resonant region.
The imaginary part of the integration comes from the path near to the pole.
Thus, we have
\begin{equation}
    \Im \int {f(r)\over J_1(r)}dr=
    \left\{\begin{array}{ll}
  -\pi f_{\rm c}/J_1'  & {\rm when}\quad (\omega-{\tilde m}\Omega)/J_1'>0 \\
   \pi f_{\rm c}/J_1'  & {\rm when}\quad (\omega-{\tilde m}\Omega)/J_1'<0.
   \end{array}
       \right. 
\label{3.6'}
\end{equation}
This can be summarized as 
\begin{equation}
   \Im  \int {f(r)\over J_1(r)}dr=
         -\pi\frac{f_{\rm c}}{\vert J_1'\vert} 
         {\rm sign} (\omega-{\tilde m}\Omega)_{\rm c},
\label{3.6}
\end{equation}
where sign$(\omega-{\tilde m}\Omega)_{\rm c}$ is the sign of  
$\omega-{\tilde m}\Omega$ at the resonant radii.

Let us now estimate the rate of work done by the resonance on the original oscillations
by using equation (\ref{3.6}).
In the case of coupling through ${\mbox{\boldmath $\xi$}}_\pm^{\rm int}$,
the work is given by equations (\ref{W+}) and (\ref{W-}). 
Since $\hat{\xi}_{r,\pm}^{\rm int}$ and
$\hat {\xi}_{\varphi,\pm}^{\rm int}$ are given by 
equations (\ref{3.1}) and (\ref{3.2}), respectively,
we have from equation (\ref{W+}) and (\ref{W-})
\begin{equation}
   W_\pm=-{\omega_0\over 2}\pi(2\pi)^{3/2}{\tilde n}!\frac{(rH\rho_{00})_{\rm c}}{\vert J_1'\vert}
           {\rm sign}(\omega-{\tilde m}\Omega)_{\rm c}
           \biggr\vert \breve{A}_{r,\pm,{\tilde n}}-i{2\Omega\over \omega-{\tilde m}\Omega}
           \breve{A}_{\varphi,\pm,{\tilde n}}
           \biggr\vert^2_{\rm c},
\label{3.7}
\end{equation}
where we have used $\int_{-\infty}^\infty {\rm exp}(z^2/2H^2){\cal H}_{\tilde n}^2(z/H)dz=
(2\pi)^{1/2}{\tilde n}!H$, and 
subscript $H$ is attached to $\omega_{\rm i}$ in order to emphasize that this is a
case of horizontal resonances.
Hence, the growth rate, $-\omega_{\rm i}$ [see equation (\ref{growth})], can be expressed as
\begin{equation}
   -\omega_{{\rm i},{\rm H}, \pm, {\tilde n}}
                =-{\pi{\tilde n}!\over 2r_{\rm c}^3
                   \vert J_1'\vert \omega_0 E_n}
       {{\rm sign}(\omega-{\tilde m}\Omega)_{\rm c}}
        \biggr\vert \breve{A}_{r,\pm,{\tilde n}}
        -i{2\Omega\over (\omega-{\tilde m}\Omega)}\breve{A}_{\varphi,\pm,{\tilde n}}
                 \biggr\vert^2_{\rm c}.
\label{5.30}
\end{equation}
It is noted that the above expression for growth rate is valid even when ${\tilde n}\not=0$.
Equations (\ref{3.7}) -- (\ref{5.30}) 
formally hold even in the case with pressure, as shown in the next section.

The vertical resonance might be unrealistic in pressure-less disks, but
we briefly summarize the formal results here, since the expression for growth rate is applicable 
even in the case of disks with pressure, as shown in the next section.
We now expand $J_2$ around $J_2=0$ as 
\begin{equation}
   J_2=J_2'\biggr[(r-r_{\rm c}) 
       -i{2(\omega-{\tilde m}\Omega)_{\rm c}\over J_2'}\omega_{\rm i}\biggr],
\label{5.32}
\end{equation}
where
\begin{equation}
    J_2'=2\biggr[{\tilde m}(\omega-{\tilde m}\Omega)\Omega{d{\rm ln}\Omega\over dr}
         +{\tilde n}\Omega_\bot^2{d{\rm ln}\Omega_\bot\over dr}\biggr]_{\rm c},
\label{5.33}
\end{equation}
and $r_{\rm c}$ is the resonant radius of $J_2=0$.
In the above two cases of the horizontal and vertical resonances, the resonant
radii are different each other.
However, the same notation, $r_{\rm c}$, is used here and hereafter without confusion.
 
After this preparation, we calculate the rate of work done on oscillations by the same
procedures as those in the case of horizontal resonance.
The results show that the growth rate is given by 
\begin{equation}
   -\omega_{{\rm i},{\rm V}, \pm, {\tilde n}}
                =-{\pi({\tilde n}-1)!\over 2r_{\rm c}^3
                   \vert J_2'\vert \omega_0 E_n}
       {{\rm sign}(\omega-{\tilde m}\Omega)_{\rm c}}
        \biggr\vert \breve{A}_{z,\pm,{\tilde n}}\biggr\vert^2_{\rm c}.
\label{5.31}
\end{equation}

\section{Resonances in Disks with Pressure}

Equations describing the intermediate oscillations [equations (\ref{4.1'}) -- 
(\ref{4.3'})
or equations (\ref{4.1''}) and (\ref{4.3'})] are now solved without assuming
$c_{\rm s}=0$ and $H=0$.
Since we are interested in behaviors of ${\breve {\mbox{\boldmath $\xi$}}}^{\rm int}$
in a region close to the
resonant point of $J_1=0$ or $J_2=0$, all the coefficients of 
${\breve {\mbox{\boldmath $\xi$}}}^{\rm int}$ 
in equations (\ref{4.1'}) -- (\ref{4.3'}) are assumed to be constant
except for $J_1$ or $J_2$.
That is, in the case of the horizontal resonance (Lindblad resonance), 
$J_2={\rm const.}$, and $J_1$ is taken as [see equations (\ref{3.4}) and (\ref{3.5})]
\begin{equation}
     J_1=J_1'x-2i(\omega-{\tilde m}\Omega)_{\rm c}\omega_{\rm i},
\label{4.1a}
\end{equation}
where $x$ is the radial distance from the resonant radius, $r_{\rm c}$,
i.e., $x=r-r_{\rm c}$.
In the case of the vertical resonance, on the other hand, $J_1={\rm const.}$, 
and $J_2$ is taken as [see equations (\ref{5.32}) and (\ref{5.33})]
\begin{equation}
     J_2=J_2'x-2i(\omega-{\tilde m}\Omega)_{\rm c}\omega_{\rm i},
\label{4.1b}
\end{equation}
where $x=r-r_{\rm c}$.
The resonant radii in the horizontal and vertical resonances are different each other,
but the same notation, $r_{\rm c}$, has been adopted here without confusion.

In disks with pressure, different from the case of pressure-less disks,
the resonant region is not narrow.
That is, the resonant region is widened by the
effects of pressure (see below), and thus the coupling terms, $\breve{A}_r$, $\breve{A}_\varphi$,
and $\breve{A}_z$, can be no longer regarded
to be spatially constant in general.
Considering these situations, 
we introduce Fourier transforms of ${\breve{\mbox{\boldmath $\xi$}}}^{\rm int}$
and $\breve{\mbox{\boldmath $A$}}$ as
\begin{equation}
    {\bar{\mbox{\boldmath{$\xi$}}}}^{\rm int}(k)=\int_{-\infty}^\infty {\rm exp}(-ikx)
          {\breve {\mbox{\boldmath $\xi$}}}^{\rm int}(x)dx,
\label{4.2a}
\end{equation}
\begin{equation}
    {\bar{\mbox{\boldmath{$A$}}}}(k)=\int_{-\infty}^\infty {\rm exp}(-ikx)
          {\breve {\mbox{\boldmath $A$}}}(x)dx.
\label{4.2b}
\end{equation}
After these preparations, 
to solve the set of equations (\ref{4.1'}) -- (\ref{4.3'}), we 
apply a method of Fourier transform adopted by Meyer-Vernet and Sicardy (1987)
in their study of resonant disk-satellite interaction. 
Here and hereafter, such subscripts as $\pm$ and ${\tilde n}$ to be attached to
${\breve {\mbox{\boldmath $\xi$}}}^{\rm int}$,
${\bar {\mbox{\boldmath $\xi$}}}^{\rm int}$, ${\breve {\mbox{\boldmath $A$}}}$,
and ${\bar {\mbox{\boldmath $A$}}}$ are sometimes 
omitted in order to avoid unnecessary complications, unless they are explicitly 
necessary.

\subsection{Horizontal Resonances (Lindblad Resonances)}

In the present case, the Fourier transform of 
equations (\ref{4.1''}) and (\ref{4.3'}) give, respectively,  
\begin{equation}
   iJ_1'{d{\bar \xi}_r^{\rm int}(k)\over dk}+i\epsilon{\bar \xi}_r^{\rm int}
       +c_{\rm s}^2k^2{\bar \xi}_r^{\rm int}+ik\Omega_\bot^2H{\bar \xi}_z^{\rm int} 
   = {\bar A}_r-i{2\Omega\over \omega-{\tilde m}\Omega}
     {\bar A}_\varphi,
\label{4.4}
\end{equation}
and
\begin{equation}
   J_2{\bar \xi}_z^{\rm int}-ik{\tilde n}\Omega_\bot^2H{\bar \xi}_r^{\rm int} 
       ={\bar A}_z,
\label{4.5}
\end{equation}
where $\epsilon\equiv -2(\omega-{\tilde m}\Omega)\omega_{\rm i}$.

Eliminating ${\bar \xi}_z^{\rm int}(k)$ from equation (\ref{4.4}) and (\ref{4.5}),
we have an inhomogeneous differential equation of ${\bar \xi}_r^{\rm int}(k)$ with respect to $k$:
\begin{equation}
     {d{\bar \xi}_r^{\rm int}(k)\over dk}+{\epsilon\over J_1'}{\bar \xi}_r^{\rm int}(k)
       + i{c_{\rm s}^2k^2\over J_1'}{(\omega-{\tilde m}\Omega)^2\over J_2}
       {\bar \xi}_r^{\rm int}(k)
       = -i{1 \over J_1'} {\bar A}(k),
\label{4.6}
\end{equation}
where ${\bar A}(k)$ is defined by
\begin{equation}
     {\bar A} \equiv {\bar A}_r-i{2\Omega\over \omega-{\tilde m}\Omega}
        {\bar A}_\varphi
        -i\frac{k\Omega_\bot^2H}{J_2}{\bar A}_z.
\label{4.7}
\end{equation}

Let us first consider the case of $\epsilon/J_1'>0$.
Equation (\ref{4.6}) is then solved to lead
\begin{equation}
   {\bar \xi}_r^{\rm int}(k)=D_{(+)}(k)
      {\rm exp}\biggr[-{\epsilon\over J_1'}k-i{c_{\rm s}^2(\omega-{\tilde m}\Omega)^2
         \over 3J_1' J_2}k^3\biggr],
\label{4.8}
\end{equation}
where
\begin{equation}
      D_{(+)}(k)=-i\frac{1}{J_1'}\int_{-\infty}^k{\bar A}(k'){\rm exp}\biggr[\frac{\epsilon}{J_1'}k'
          +i\frac{c_{\rm s}^2(\omega-{\tilde m}\Omega)^2}{3J_1'J_2}k'^3\biggr]dk'.
\label{4.8a}
\end{equation}
In determining the integration range of equation (\ref{4.8a}), we have used a boundary 
condition allowing
${\bar \xi}_r^{\rm int}(k)\rightarrow 0$ for $k\rightarrow -\infty$.
This boundary condition is required, since ${\bar \xi}_r^{\rm int}(k)$ is 
the Fourier transform of ${\breve \xi}_r^{\rm int}(x)$.

By using the inverse Fourier transform:
\begin{equation}
   {\breve \xi}_r^{\rm int}(x)={1\over 2\pi}\int_{-\infty}^\infty 
       {\bar \xi}_r^{\rm int}(k){\rm exp}(ikx) dk,
\label{4.10}
\end{equation}
we can express ${\breve \xi}_r^{\rm int}(x)$ in an integration form as
\begin{equation}
      {\breve \xi}_r^{\rm int}(x) = \frac{1}{2\pi}
          \int_{-\infty}^\infty D_{(+)}(k){\rm exp}\biggr[-\frac{\epsilon}{J_1'}k
              +i\biggr(kx-\frac{1}{3}\alpha_{\rm P}^3k^3\biggr)\biggr]dk,
\label{xi-r-again}
\end{equation}
where $\alpha_{\rm P}$ is defined by
\begin{equation}
    \alpha_{\rm P}^3(x)\equiv \frac{c_{\rm s}^2(\omega-{\tilde m}\Omega)^2}
            {J_1'J_2}.
\label{alpha-p}
\end{equation}
Since  ${\breve \xi}_\varphi^{\rm int}(x)$ and ${\breve {\xi_z^{\rm int}}}(x)$ are related to
${\breve \xi}_r^{\rm int}$ by
[see equations (\ref{4.2'}) and (\ref{4.3'})]
\begin{equation}
   {\breve \xi}_\varphi^{\rm int}(x)=i\frac{2\Omega}{\omega-{\tilde m}\Omega}
   {\breve \xi}_r^{\rm int}-\frac{{\breve A}_\varphi}{(\omega-{\tilde m}\Omega)^2},
   \quad 
   {\breve \xi}_z^{\rm int}=\frac{1}{J_2}\biggr({\tilde n}\Omega_\bot^2H
            \frac{d{\breve \xi}_r^{\rm int}}{dr}+{\breve A}_z\biggr),
\label{xi-varphi-again}
\end{equation}
the work integral, $W_{{\rm H},+}$, in the case of the coupling through 
$\breve {\mbox{\boldmath $\xi$}}^{\rm int}_{+,{\tilde n}}$,
is written as, using equation (\ref{W+}),
\begin{eqnarray}
   W_{{\rm H},+}&&=\frac{\omega_0}{2}\Im \int dx\int_{-\infty}^\infty dk
      \frac{1}{2\pi}\rho_{00}[(2\pi)^{3/2}{\tilde n}!rH]
       D_{(+)}(k) {\rm exp}\biggr[-\frac{\epsilon}{J_1'}k+i\biggr(kx-\frac{1}{3}\alpha_{\rm P}^3k^3\biggr)
       \biggr] \nonumber \\
       &&\times\biggr[{\breve A}_{r,+}^*(x)+i\frac{2\Omega}{\omega-{\tilde m}
             \Omega}{\breve A}_{\varphi,+}^*(x)
         +\frac{1}{J_2}ik\Omega_\bot^2H{\breve A}_{z,+}^*(x)\biggr],
\label{W+2}
\end{eqnarray}
where real parts on the right-hand side of equation (\ref{W+2}) have been omitted.

In the case of $\epsilon/J_1'<0$, the expression for $\bar{\xi}_r^{\rm int}(k)$ is slightly
changed from equation (\ref{4.8}) as
\begin{equation}
   {\bar \xi}_r^{\rm int}(k)=D_{(-)}(k)
      {\rm exp}\biggr[-{\epsilon\over J_1'}k-i{c_{\rm s}^2(\omega-{\tilde m}\Omega)^2
         \over 3J_1' J_2}k^3\biggr],
\label{4.8'}
\end{equation}
where $D_{(-)}(k)$ is given by
\begin{equation}
      D_{(-)}(k)=-i\frac{1}{J_1'}\int_{\infty}^k{\bar A}(k'){\rm exp}\biggr[\frac{\epsilon}{J_1'}k'+
         \frac{c_{\rm s}^2(\omega-{\tilde m}\Omega)^2}{3J_1'J_2}k'^3\biggr]dk',
\label{4.8a'}
\end{equation}
where $\bar {A}(k)$ is defined by equation (\ref{4.7}).
The integration range in equation (\ref{4.8a'}) is changed from equation (\ref{4.8a}) 
so that the boundary condition $\bar{\xi}_r^{\rm int}=0$ is satisfied at $k=\infty$.
By performing the same procedures as those in the case of $\epsilon/J_1'>0$, we can derive
an expression for $W_{{\rm H},+}$.
An expression for $W_{{\rm H},+}$ in this case is the same as equation (\ref{W+2}), except that
$D_{(+)}$ is now changed to $D_{(-)}$, i.e., 
\begin{eqnarray}
   W_{{\rm H},+}&&=\frac{\omega_0}{2}\Im \int dx\int_{-\infty}^\infty dk
      \frac{1}{2\pi}\rho_{00}[(2\pi)^{3/2}{\tilde n}!rH]
       D_{(-)}(k) {\rm exp}\biggr[-\frac{\epsilon}{J_1'}k+i\biggr(kx-\frac{1}{3}\alpha_{\rm P}^3k^3\biggr)
       \biggr] \nonumber \\
       &&\times\biggr[{\breve A}_{r,+}^*(x)+i\frac{2\Omega}{\omega-{\tilde m}
             \Omega}{\breve A}_{\varphi,+}^*(x)
         +\frac{1}{J_2}ik\Omega_\bot^2H{\breve A}_{z,+}^*(x)\biggr],
\label{W+2a}
\end{eqnarray}

Expressions of the work integral in the case of the coupling through 
$\breve{\mbox{\boldmath $\xi$}}_{-, {\tilde n}}^{\rm int}$ can be derived by similar 
procedures as those in the case of the coupling through  
$\breve{\mbox{\boldmath $\xi$}}_{+, {\tilde n}}^{\rm int}$.
In this case, the subscript + attached to $A$'s in equations (\ref{W+2}) and (\ref{W+2a})
is changed to $-$ and ${\tilde m}=m-1$.

\subsubsection {The case of constant coupling terms}

First, we consider the case where the coupling terms, $\breve{A}_r$,
$\breve{A}_\varphi$, and $\breve{A}_z$,  can be regarded to be 
spatially constant in the resonant region.
(This is really realized in a particular case, as mentioned in the next section.)
An another purpose of examining the case of constant coupling terms is to have a rough image on 
the width of resonant region in disks with pressure.

If ${\breve A}_r$, ${\breve A}_\varphi$, and ${\breve A}_z$ are spatially constant, 
their Fourier transforms give
\begin{equation}
  {\bar A}_r(k)=2\pi {\breve A}_r\delta(k), \quad
  {\breve A}_\varphi=2\pi{\breve A}_\varphi\delta(k), \quad
  {\bar A}_z(k)=2\pi {\breve A}_z\delta(k).
\label{constant}
\end{equation}
In this case, $D_{(+)}(k)$ given by equation (\ref{4.8a}) is reduced simply to
\begin{equation}
    D_{(+)}(k)=-i\frac{2\pi}{J_1'}H_+(k)
      \biggr[{\breve A}_r-i\frac{2\Omega}{\omega-{\tilde m}\Omega}{\breve A}_\varphi\biggr],
\label{D}
\end{equation}
where $H_+(k)$ is the unit-step function, i.e., $H_+(k)=0$ for $k<0$, while $H(k)=1$ for
$k>0$.
Since $D_{(+)}(k)$ does not involve $x$, the integration with respect to $x$ in equation (\ref{W+2})
leads to $2\pi\delta(k)$.
Hence, performing the integration with respect to $k$ 
in equation (\ref{W+2}), we have
\begin{equation}
    W_{{\rm H},+}=-\frac{\omega_0}{2}\pi(2\pi)^{3/2}{\tilde n}!
           \frac{(rH\rho_{00})_{\rm c}}{J_1'}
           \biggr\vert {\breve A}_r-i\frac{2\Omega}{\omega-{\tilde m}\Omega)}{\breve A}_\varphi
           \biggr\vert^2,
\label{W+3}
\end{equation} 
in the case of $\epsilon/J_1'>0$, where 
${\tilde m}=m+1$ and ${\tilde n}=n-1,\ {\rm or}\ n,\ {\rm or}\ n+1$, depending on the type of
couplings and the form of deformation.

In the case of $\epsilon/J_1'<0$, ${\bar \xi}_r^{\rm int}(k)$ is given by
equation (\ref{4.8'}).
By performing the same procedures as those in the case of $\epsilon/J_1'>0$, we can derive
an expression for $W_{{\rm H},+}$.
The final expression for $W_{{\rm H},+}$ in this case has the opposite sign from equation (\ref{W+3}).

Inequality $\epsilon/J_1'>0$ implies that $(\omega-{\tilde m}\Omega)/J_1'>0$,
since $-\omega_{\rm i}$ should be taken tentatively to be negative as mentioned before,
while inequality $\epsilon/J_1<0$ means $(\omega-{\tilde m}\Omega)/J_1'>0$.
Considering this difference, we can summarize $W_{{\rm H},+}$ 
in the above two cases of $\epsilon/J_1'>0$ and $\epsilon/J_1'<0$ as 
\begin{equation}
   W_{{\rm H},+,{\tilde n}}=-{\omega_0\over 2}\pi(2\pi)^{3/2}{\tilde n}!
          \frac{(rH\rho_{00})_{\rm c}}{\vert J_1'\vert}
       {\rm sign}(\omega-{\tilde m}\Omega)_{\rm c}\biggr\vert {\breve A}_{r,+,{\tilde n}}
    -i{2\Omega\over \omega-{\tilde m}\Omega}{\breve A}_{\varphi,,+,{\tilde n}}
        \biggr\vert^2,
\label{4.17}
\end{equation}
where ${\tilde m}=m+1$.
 
In the case of coupling through ${{\mbox{\boldmath $\xi$}}}_-^{\rm int}$,
we have, after the similar processes as the above,
\begin{equation}
   W_{{\rm H},-,{\tilde n}}=-{\omega_0\over 2}\pi(2\pi)^{3/2}{\tilde n}!
       \frac{(rH\rho_{00})_{\rm c}}{\vert J_1'\vert}
       {\rm sign}(\omega-{\tilde m}\Omega)_{\rm c}
       \biggr\vert {\breve A}_{r,-,{\tilde n}}
    -i{2\Omega\over \omega-{\tilde m}\Omega}
       {\breve A}_{\varphi, -, {\tilde n}}\biggr\vert^2,
\label{4.18}
\end{equation}
where ${\tilde m}=m-1$.

Finally, from $W_{{\rm H},\pm,{\tilde n}}$ given above and the wave energy, $E$,
given by equation (\ref{2.12}), we find that the growth rate, equation (\ref{5.28}),
is written as
\begin{equation}
     -\omega_{{\rm i},{\rm H},\pm,{\tilde n}}=
          -{\pi{\tilde n}! \over 2r_{\rm c}^3\vert J_1'\vert \omega_0 E_n}
           {\rm sign}(\omega-{\tilde m}\Omega)_{\rm c}
          \biggr\vert {\breve A}_{r,\pm,{\tilde n}}
    -i{2\Omega\over \omega-{\tilde m}\Omega}
       {\breve A}_{\varphi, \pm, {\tilde n}}\biggr\vert^2,  
\label{4.19}
\end{equation}
which is identical to equation (\ref{5.30}), and also
equal to the growth rate derived in the previous paper (Kato 2008a),
as will be mentioned  at the end of this section.

\subsubsection{The case of radially changing coupling terms}

The arguments in subsection 4.1.1 show that in the case where ${\breve A}_r$, ${\breve A}_\varphi$,
and ${\breve A}_z$ are radially constant, the radial structure of resonant region is 
determined by [see equation (\ref{W+2}) and notice that $D_{(+)}(k)$ has the
step-function]
\begin{equation}
    \Re\int_0^\infty dk\ {\rm exp}\biggr[-\frac{\epsilon}{J_1'}k
        +i\biggr(kx-\frac{1}{3}\alpha_{\rm P}^3k^3\biggr)\biggr],
\end{equation}
where the term with $\epsilon$ can be practically neglected.
The $x$-dependence of this integral has been examined by Meyer-Vernet and Sicardy (1987)
[see equation (41) and Fig, 5 in their paper].
They show that the half width of the resonant region is $\sim \alpha_{\rm P}$, which
is on the order of $(rH^2)^{1/3}$, i.e., $r(H/r)^{2/3}$, in the present case.

In general, ${\breve A}_r$, ${\breve A}_\varphi$, and ${\breve A}_z$ 
radially oscillate with the wavelength determined by the product of $\breve {\mbox{\boldmath $\xi$}}$
and $\breve {\mbox{\boldmath $\xi$}}^{\rm W}$.
Hence, in some cases the scale-length of the variation of ${\breve A}$'s will be shorter 
than $r(H/r)^{2/3}$, 
although it depends on where the resonance occurs (see the next section).
To examine general cases, numerical calculations are needed.
However, in order to have a rough image of the effects of spatial variation of 
${\breve A}_r$, ${\breve A}_\varphi$, and ${\breve A}_z$
on the work integral $W_{{\rm H},\pm,{\tilde n}}$, we consider here a limiting case where
${\breve A}_r$, ${\breve A}_\varphi$, and ${\breve A}_z$ oscillate periodically in 
the radial direction.
That is, we assume 
\begin{equation}
    {\breve A}_r(x)={\breve A}_r^{(0)}{\rm exp}(ik_0x), \quad
    {\breve A}_\varphi(x)={\breve A}_\varphi^{(0)}{\rm exp}(ik_0x), \quad
    {\breve A}_z(x)={\breve A}_z^{(0)}{\rm exp}(ik_0x),
\label{periodic}
\end{equation}
where ${\breve A}^{(0)}$'s and $k_0$ are the amplitudes and wavenumber of the 
spatial variations.
Then, the Fourier transform defined by equation (\ref{4.2b}) show that
\begin{equation}
    {\bar A}_r(k)=2\pi {\breve A}_r^{(0)}\delta (k-k_0),\quad
    {\bar A}_\varphi(k)=2\pi {\breve A}_\varphi^{(0)}\delta (k-k_0), \quad
    {\bar A}_z(k)=2\pi {\breve A}_z^{(0)}\delta (k-k_0).
\label{}
\end{equation}

Let us consider the case of $\epsilon/J_1'>0$.
In this case, $D_{(+)}(k)$ given by equation (\ref{4.8a}) is reduced to [cf., equation (\ref{D})]
\begin{equation}
     D_{(+)}(k)=-i{2\pi\over J_1'}H_+(k-k_0)
         \biggr[{\breve A}_r^{(0)}-i\frac{2\Omega}{\omega-{\tilde m}\Omega}
         {\breve A}_\varphi^{(0)}
             -i\frac{k_0\Omega_\bot^2H}{J_2}
         {\breve A}_z^{(0)}\biggr]
             {\rm exp}\biggr[\frac{\epsilon}{J_1'}k_0+i\frac{\alpha_{\rm P}^3}{3}k_0^3\biggr].
\label{newD}
\end{equation}

Since ${\breve A}_r$, ${\breve A}_\varphi$, and ${\breve A}_z$ are given by equation (\ref{periodic}),
performance of integration with respect to $x$ in equation (\ref{W+2}) gives 
$2\pi\delta(k-k_0)$.
Next, we perform the integration with respect to $k$.
Since $D_{(+)}(k)$ is now given by equation (\ref{newD}), we have
\begin{equation}
    W_{{\rm H},+}=-\frac{\omega_0}{2}\pi(2\pi)^{3/2}{\tilde n}!
           \frac{(rH\rho_{00})_{\rm c}}{J_1'}
      \biggr\vert {\breve A}_r^{(0)}-i\frac{2\Omega}{\omega-{\tilde m}\Omega}
              {\breve A}_\varphi^{(0)}-i\frac{k_0\Omega_\bot^2H}{J_2}{\breve A}_z^{(0)}\biggr\vert^2.
\label{W+4}
\end{equation}
 
Similar arguments can be made in the case of $\epsilon/J_1'<0$.
In this case, $W_{{\rm H},+}$ has the sign opposite to equation (\ref{W+4}).
Expressions for $W_{{\rm H},+}$ in two cases of $\epsilon/J_1'>0$ and $\epsilon/J_1'<0$ 
can be combined into a single form as before.
Similarly, we can consider the case where the coupling occurs through 
$\breve {\mbox{\boldmath $\xi$}}_-^{\rm int}$.
All of these results can be summarized as
\begin{eqnarray}
    W_{{\rm H},\pm}&&=-\frac{\omega_0}{2}\pi(2\pi)^{3/2}{\tilde n}!
           \frac{(rH\rho_{00})_{\rm c}}{\vert J_1'\vert} 
           {\rm sign}(\omega-{\tilde m}\Omega)_{\rm c} \nonumber \\
           &&\times\biggr\vert {\breve A}_{r,\pm,{\tilde n}}^{(0)}-
                 i\frac{2\Omega}{\omega-{\tilde m}\Omega}
              {\breve A}_{\varphi,\pm,{\tilde n}}^{(0)}
              -i\frac{k_0\Omega_\bot^2H}{J_2}{\breve A}_{z,\pm,{\tilde n}}^{(0)}\biggr\vert^2.
\label{W+5}
\end{eqnarray} 
Hence, the growth rate is written as
\begin{eqnarray}
     -\omega_{{\rm i},{\rm H},\pm,{\tilde n}}=&&
          -{\pi{\tilde n}! \over 2r_{\rm c}^3\vert J_1'\vert \omega_0 E_n}
           {\rm sign}(\omega-{\tilde m}\Omega)_{\rm c}  \nonumber \\  
          &&\times \biggr\vert {\breve A}^{(0)}_{r,\pm,{\tilde n}}
    -i{2\Omega\over \omega-{\tilde m}\Omega}
       {\breve A}^{(0)}_{\varphi, \pm, {\tilde n}}
       -i\frac{k_0\Omega_\bot^2H}{J_2}{\breve A}_{z,\pm,{\tilde n}}^{(0)}
       \biggr\vert^2.
\label{4.19'}
\end{eqnarray}

\subsection{Vertical Resonances}

In the case of the vertical resonance, from equations (\ref{4.1''}) and (\ref{4.3'}),
we have, respectively, 
\begin{equation}
   J_1{\bar \xi}_r^{\rm int}+k^2c_{\rm s}^2{\bar \xi}_r^{\rm int}
        +ik\Omega_\bot^2H{\bar \xi}_z^{\rm int}={\bar A}_r-i\frac{2\Omega}{\omega-{\tilde m}\Omega}
             \bar{A}_\varphi,
\label{v1}
\end{equation}
and
\begin{equation}
     iJ_2'{d\xi_z^{\rm int}\over dk}+i\epsilon{\bar \xi}_z^{\rm int}
       -ik{\tilde n}\Omega_\bot^2H{\bar \xi}_r^{\rm int}
       = {\bar A}_z.
\label{v2}
\end{equation}
Eliminating ${\bar \xi}_r^{\rm int}$ from the above two equations, we have
an inhomogeneous differential equation with respect to ${\bar \xi}_z^{\rm int}$ as
\begin{equation}
   {d\bar {\xi}_z^{\rm int}\over dk}+{\epsilon\over J_2'}\bar {\xi}_z^{\rm int}
         +i\alpha_{\rm P,V}k^2\bar {\xi}_z
            =-i\frac{1}{J_2'}\biggr[\frac{ik{\tilde n}\Omega_\bot^2 H}{J_1}
             \biggr({\bar A}_r-i\frac{2\Omega}{\omega-{\tilde m}\Omega}{\bar A}_\varphi\biggr)
                +{\bar A}_z\biggr],
\label{v4}
\end{equation}
where 
\begin{equation}
   \alpha_{\rm P,V}^3= \frac{{\tilde n}c_{\rm s}^2\Omega_\bot^2}{J_2'J_1}.
\label{v5}
\end{equation}
In deriving equation (\ref{v4}), in order to avoid unnecessary complications, we have neglected 
$k^2c_{\rm s}^2$ compared with $J_1$, since the magnitude 
of $kH$ contributing to the present formulations is smaller than unity, i.e., $kH\ll 1$.

In the case of $\epsilon/J_2'>0$, equation (\ref{v4}) is solved as
\begin{equation}
   {\bar\xi}^{\rm int}_z(k)=D_{{\rm V},(+)}(k){\rm exp}\biggr[-\frac{\epsilon}{J_2'}k
        -i{1\over 3}\alpha_{\rm P,V}^3k^3\biggr],
\label{v6}
\end{equation}
where the coefficient $D_{{\rm V},(+)}(k)$ is given by
\begin{eqnarray}
     D_{{\rm V},(+)}(k)=&&-i{1\over J_2}\int_{-\infty}^k 
     \biggr[\frac{ik'{\tilde n}\Omega_\bot^2 H}{J_1}
             \biggr({\bar A}_r-i\frac{2\Omega}{\omega-{\tilde m}\Omega}{\bar A}_\varphi\biggr)
                +{\bar A}_z\biggr] \nonumber \\
       &&\times {\rm exp}
       \biggr[\frac{\epsilon}{J_2'}k'+i{1\over 3}\alpha_{\rm P,V}^3k'^3\biggr]dk'.
\label{v7}
\end{eqnarray}
By using ${\bar \xi}_z^{\rm int}(k)$ given above, we can write 
${\bar \xi}_r^{\rm int}$  and ${\bar \xi}_\varphi^{\rm int}$ as
[see equations (\ref{4.1''}) and (\ref{4.2'})],
\begin{equation}
    {\bar\xi}_r^{\rm int}=-i\frac{k\Omega_\bot^2H}{J_1}{\bar\xi}_z^{\rm int}
        +{1\over J_1}\biggr({\bar A}_r-i\frac{2\Omega}{\omega-{\tilde m}\Omega}{\bar A}_\varphi\biggr)
\label{v8}
\end{equation}
and
\begin{equation}
    {\bar \xi}_\varphi^{\rm int}=\frac{2\Omega}{\omega-{\tilde m}\Omega}
            \frac{k\Omega_\bot^2H}{J_1}{\bar \xi}_z^{\rm int}
            +i{1\over J_1}\frac{2\Omega}{\omega-{\tilde m}\Omega}
           \biggr({\bar A}_r-i\frac{2\Omega}{\omega-{\tilde m}\Omega}{\bar A}_\varphi\biggr).
\label{v9}
\end{equation}
Substitution of these relations into the work integral (\ref{W+}) gives
\begin{eqnarray}
   W_{{\rm V},+}&&=\frac{\omega_0}{2}\Im \int dx\int_{-\infty}^\infty dk
      \frac{1}{2\pi}\rho_{00}[(2\pi)^{3/2}{\tilde n}!rH]
      D_{{\rm V},(+)}(k) {\rm exp}\biggr[-\frac{\epsilon}{J_2'}k+
       i\biggr(kx-\frac{1}{3}\alpha_{\rm P,V}^3k^3\biggr)\biggr]  \nonumber \\
     &&\times
       \biggr[-i\frac{ik\Omega_\bot^2 H}{J_1}
             \biggr({\bar A}_r^*+i\frac{2\Omega}{\omega-{\tilde m}\Omega}{\bar A}_\varphi^*\biggr)
                +\frac{1}{{\tilde n}}{\bar A}_z^*\biggr].      
\label{v10}
\end{eqnarray}

In the case of $\epsilon/J_2'<0$, $W_{{\rm V},+}$ has a similar form, but
we should use, instead of $D_{{\rm V},(+)}(k)$,
$D_{{\rm V},(-)}(k)$, which is given by
\begin{eqnarray}
     D_{{\rm V},(-)}(k)=&&-i{1\over J_2}\int_{\infty}^k 
     \biggr[\frac{ik'{\tilde n}\Omega_\bot^2 H}{J_1}
             \biggr({\bar A}_r-i\frac{2\Omega}{\omega-{\tilde m}\Omega)}{\bar A}_\varphi\biggr)
                +{\bar A}_z\biggr]  \nonumber \\
          &&\times {\rm exp}
       \biggr[\frac{\epsilon}{J_2'}k'+i{1\over 3}\alpha_{\rm P,V}^3k'^3\biggr]dk'.
\label{v7'}
\end{eqnarray}

As in the subsection 4.1.1, we consider two cases where i) ${\breve A}_z$ is constant 
in the resonant region and
ii) ${\breve A}_z$ is spatially periodic as ${\breve A}_z^{(0)}{\rm exp}(ik_0x)$.
In the former case, summing the two cases of $\epsilon/J_2' > 0$ and
$\epsilon/J_2'$, we have 
\begin{equation}
   W_{{\rm V},+,{\tilde n}}=-{\omega_0 \over 2}\pi(2\pi)^{3/2}({\tilde n}-1)!
     \frac{(rH\rho_{00})_{\rm c}}{\vert J_2'\vert}
         {\rm sign}(\omega-{\tilde m}\Omega)_{\rm c}   
          \vert{\breve A}_{z,+,{\tilde n}}\vert^2.  
\label{v11a}
\end{equation}   
and in the latter case we have 
\begin{eqnarray}
   W_{{\rm V},+,{\tilde n}}=&&-{\omega_0 \over 2}\pi(2\pi)^{3/2}({\tilde n}-1)!
     \frac{(rH\rho_{00})_{\rm c}}{\vert J_2'\vert}
         {\rm sign}(\omega-{\tilde m}\Omega)_{\rm c}   \nonumber \\
          &&\times\biggr\vert i\frac{{\tilde n}k_0\Omega_\bot^2H}{J_1}
          \biggr(\breve{A}_{r,+,{\tilde n}}^{(0)}
          -i\frac{2\Omega}{\omega-{\tilde m}}\breve{A}_{\varphi,+,{\tilde n}}^{(0)}\biggr)
          + {\breve A}_{z,+,{\tilde n}}\biggr\vert^2.  
\label{v11}
\end{eqnarray}     

In the coupling through $\breve {\mbox{\boldmath $\xi$}}_-$, the work integral, 
$W_{{\rm V},-,{\tilde n}}$,
has the same expression as equations (\ref{v11a}) and (\ref{v11}), except that  
the subscript $+$ is now changed to $-$.

After the above considerations, we summarize the growth rate, 
$\omega_{{\rm i},{\rm V},\pm,{\tilde n}}$, in the former case of $\breve{A}_z$ being constant as 
\begin{equation}
     -\omega_{{\rm i},{\rm V},\pm,{\tilde n}}=
          -{\pi({\tilde n}-1)! \over 2r_{\rm c}^3\vert J_2'\vert \omega_0 E_n}
           {\rm sign}(\omega-{\tilde m}\Omega)_{\rm c} 
          \biggr\vert {\breve A}_{z,\pm,{\tilde n}}\biggr\vert^2.
\label{v12}
\end{equation}
In the case where ${\breve A}_z$ changes as ${\breve A}_z={\breve A}_z^{(0)}{\rm exp}
(ik_0x)$, we have
\begin{eqnarray}
     -\omega_{\rm i,{\rm V},\pm,{\tilde n}}=&&
          -{\pi({\tilde n}-1)! \over 2r_{\rm c}^3\vert J_2'\vert \omega_0 E_n}
           {\rm sign}(\omega-{\tilde m}\Omega)_{\rm c}    \nonumber \\
          && \times\biggr\vert i\frac{{\tilde n}k_0\Omega_\bot^2H}{J_1}
          \biggr(\breve{A}_{r,+,{\tilde n}}^{(0)}
          -i\frac{2\Omega}{\omega-{\tilde m}}\breve{A}_{\varphi,+,{\tilde n}}^{(0)}\biggr)
          +{\breve A}_{z,\pm,{\tilde n}}^{(0)}\biggr\vert^2.
\label{v12'}
\end{eqnarray}

Finally, we note here that the growth rates obtained in the previous papers
are the same as those obtained in subsection 4.1.1, although the final expression in the 
previous papers are different from those in subsection 4.1.1.
In the case of horizontal resonance, for example, $G_{{\rm H},\pm}$ in Kato (2008a)
is equal to $r_{\rm c}J_1'/2$ in the present paper.
Furthermore, if ${\breve \xi}_\varphi^{\rm int}$ and ${\breve \xi}_z^{\rm int}$
are eliminated from equations (46) -- (48) in Kato (2008a), we have
\begin{equation}
     {\breve \xi}_r^{\rm int}={J_2\over D}\biggr({\breve A}_r
      -i{2\Omega\over \omega-{\tilde m}\Omega}{\breve A}_\varphi\biggr),
\label{*}
\end{equation}
where $D$ is $J_1J_2-c_{\rm s}^2k^2(\omega-{\tilde m}\Omega)^2$.
This expression for ${\breve \xi}_r^{\rm int}$ shows that ${\breve \zeta}_r$
defined by equation (52) in Kato (2008a) is written as
\begin{equation}
   {\breve \zeta}_r=J_2\biggr({\breve A}_r
      -i{2\Omega\over \omega-{\tilde m}\Omega}{\breve A}_\varphi\biggr).
\label{**}
\end{equation}
Then, substituting $G_{\rm H}=r_{\rm c}J_1'/2$ and equation (\ref{**}) with 
$(J_2)_{\rm c}=(-\kappa^2+{\tilde n}\Omega_\bot^2)_{\rm c}$ into equation (70) in
Kato (2008a), we obtain an expression for growth rate in the case of horizontal
resonances, which is identical with equation (\ref{4.19}) in subsection 4.1.1.

Similar arguments easily show that the growth rate given by Kato (2008a) for
vertical resonances, i.e., equation (71) in Kato (2008a), is
identical with that in section 4.2, i.e., equation (\ref{v11}).

\section{Discussion}

In this paper we have derived a stability criterion on resonant excitation of
disk oscillations in a deformed disk.
Deformation of disks that can lead to excitation of disk oscillations are
a warp and an eccentric deformation of disk plane,
i.e., $m^{\rm W}=1$, although this is not 
discussed in this paper (see Kato 2004, 2008a).
In previous papers (Kato 2004, 2008a), the above resonant excitation problem 
was examined and a condition of excitation of disk oscillations was derived.
From mathematical point of view, however, the treatment of disk oscillations around 
the resonant point was not suitable in the previous papers.
In this paper, we corrected the inadequate points.
The results of the correction show that the stability condition and the growth rate
derived in the previous papers are kept qualitatively unchanged.
In this paper, however, the growth rate of oscillations is written in a form applicable
more easily to numerical calculations of growth rate.

In disks with pressure, the resonant region is widened by pressure effects.
Based on arguments by Meyer-Vernet and Sicardy (1987), we showed in section 4 that
the half-width of the resonant region is $\sim r(H/r)^{2/3}$ (or $\sim H(r/H)^{1/3}$),
which is shorter than $r$ but longer than $H$.
The coupling terms comes from nonlinear products of original oscillation,
$\breve{\mbox{\boldmath $\xi$}}$, and disk deformation,
$\breve{\mbox{\boldmath $\xi$}}^{\rm W}$.
If the disk deformation is assumed to be global, the scale of spatial variation 
of the coupling terms is mainly governed by the spatial behavior of the
original oscillation, $\breve{\mbox{\boldmath $\xi$}}$.
Hence, in general, the width of resonant region is comparable with the wavelength 
of the original oscillations and may not be narrow enough to
regard the coupling terms $\breve{A}$'s as constant in the region.
Hence, in section 4, we examined two limiting cases where i) the coupling terms are
constant in the resonant region and ii) they vary sinusoidally in the region.
Analyses in section 4 show that in these two limiting cases there are no
essential differences in growth rate, especially no difference in 
the criterion of stability.
To examine what happens in intermediate cases of the above two limiting ones,
however, numerical calculations will be needed.

It is noted that the case where the coupling terms, $\breve{\mbox{\boldmath $A$}}$,
are roughly constant in the resonant region is realized when resonance occurs
near to the boundary between the propagation and evanescent regions of the
the original oscillations.
This is because near to the boundary, the radial wavelength of the
oscillations is long.
Applications of the present disk oscillation model to high-frequency QPOs by
Kato and Fukue (2006) and Kato (2008a) are concerned in such cases.
The another limiting case where the coupling terms vary in the radial direction 
will be realized when  resonance occurs in the propagation region far apart from 
the boundary between the propagation and evanescent regions.
 
There are two kinds of resonances, i.e., horizontal and vertical resonances.
In both resonances, the condition of excitation of oscillations is that the
wave energy of the original oscillation, $E_n$, and $\omega-{\tilde m}\Omega$ at 
the resonant radius have the opposite signs, i.e., $E_n\cdot 
{\rm sign}(\omega-{\tilde m}\Omega)_{\rm c}<0$.
For the resonance to occur efficiently, the resonant radius must be within the region
where both original and intermediate oscillations exists predominantly.
Hence, it will be reasonable to suppose that the sign of $(\omega-{\tilde m}\Omega)_{\rm c}$
roughly represents the sign of the energy of the intermediate oscillation.
[Notice that the sign of wave energy of the intermediate oscillation is determined by 
the sign of $\omega-{\tilde m}\Omega$ in the region where the wave exists predominantly,
see equations (\ref{5.28}) and (\ref{5.29})].
In other words, the condition of $E_n\cdot{\rm sign}(\omega-{\tilde m}\Omega)_{\rm c}<0$
is roughly equal to the condition that the original and intermediate oscillations
have energies of the opposite signs.
This instability criterion thus can be understand as a result of energy exchange 
between the original and intermediate oscillations at the resonant point
(Ferreira and Ogilvie 2008).

The resonant instability condition, i.e., $E_n\cdot{\rm sign}(\omega-{\tilde m}\Omega)_{\rm c}
<0$, however, is not exactly equal to the condition of both oscillaions
(original and intermediate one) having the opposite signs.
This suggests that the energy exchange at the resonant point is
not only between the above two oscillations, but the disk rotation also plays
a part and gains or losses its energy.
In the case of $E_n\cdot{\rm sign}(\omega-{\tilde m}\Omega)_{\rm c}>0$, the original oscillation 
is damped.
Roughly speaking, in this case, both oscillations (original and intermediate ones)
have the same signs in their wave energy.
Hence, in this case, the main energy flow will be between disk rotation and the
oscillations so that the oscillations are damped.
It is noticed that in vertical resonance we always have 
$E_n\cdot{\rm sign}(\omega-{\tilde m}\Omega)_{\rm c}>0$ in the case where
a resonant radius can appear in disk plane and the oscillations are always damped 
(see Kato 2004, 2008a).

Finally, it is noted that the present resonant excitation mechanism
has been applied to quasi-periodic oscillations observed in black-hole and neutron-star 
X-ray binaries, for example, by Kato and Fukue (2006), Kato (2008a) and Kato
et al. (2008).

\bigskip
The author thanks the referee for very careful reading of the manuscript with 
valuable comments, especially for pointing out widening of resonant region by
pressure.

\bigskip
\leftskip=20pt
\parindent=-20pt
\par
{\bf References}
\par
Ferreira, B.T. and Ogilvie, G.I. 2008, arXiv:0803.4123 \par
Kato, S. 2001, PASJ, 53, 1\par 
Kato, S. 2004, PASJ, 56, 905\par
Kato, S. 2008a, PASJ, 60, 111 \par
Kato, S. 2008b, PASJ, 60, No. 4 in press (arXiv:0803.2384) \par
Kato, S., Fukue, J., \& Mineshige, S. 1998, Black-Hole Accretion Disks 
  (Kyoto: Kyoto University Press)\par
Kato, S. and Fukue, J. 2006, PASJ, 58, 909 \par  
Kato, S., Fukue, J., \& Mineshige, S. 2008, Black-Hole Accretion Disks 
  -- Toward a New paradigm -- (Kyoto: Kyoto University Press)\par
Lynden-Bell, D. and Ostriker, J.P. 1967, MNRAS, 136, 293  \par
Meyer-Vernet, N. and Sicardy, B. 1987, Icarus, 69, 157 \par

\end{document}